\documentclass[aip,apl,preprint]{revtex4-1}

\usepackage{graphicx}

\begin{document}

\title{Coupling of erbium dopants to yttrium orthosilicate photonic crystal cavities for on-chip optical quantum memories}

\author{Evan Miyazono}
\affiliation{T. J. Watson Laboratory of Applied Physics, California Institute of Technology, 1200 E California Blvd, Pasadena, CA, 91125, USA}
\author{Tian Zhong}
\affiliation{T. J. Watson Laboratory of Applied Physics, California Institute of Technology, 1200 E California Blvd, Pasadena, CA, 91125, USA}
\author{Ioana Craiciu}
\affiliation{T. J. Watson Laboratory of Applied Physics, California Institute of Technology, 1200 E California Blvd, Pasadena, CA, 91125, USA}
\author{Jonathan M. Kindem}
\affiliation{T. J. Watson Laboratory of Applied Physics, California Institute of Technology, 1200 E California Blvd, Pasadena, CA, 91125, USA}
\author{Andrei Faraon}
\email{faraon@caltech.edu}
\affiliation{T. J. Watson Laboratory of Applied Physics, California Institute of Technology, 1200 E California Blvd, Pasadena, CA, 91125, USA}


\begin{abstract} 
Erbium dopants in crystals exhibit highly coherent optical transitions well suited for solid-state optical quantum memories operating in the telecom band. Here we demonstrate coupling of erbium dopant ions in yttrium orthosilicate to a photonic crystal cavity fabricated directly in the host crystal using focused ion beam milling. The coupling leads to reduction of the photoluminescence lifetime and enhancement of the optical depth in microns-long devices, which will enable on-chip quantum memories. 
\end{abstract}


\maketitle
Optical quantum networks\cite{Kimble2008} are currently investigated for applications that require distribution of quantum entanglement over long distances. Optical quantum memories are essential network components that enable high fidelity storage and retrieval of photonic states.~\cite{Northup2014} However, for practical applications, solid-state components are desirable. Rare-earth doped crystals are state-of-the-art materials that have been used for high performance optical memories.~\cite{Lvovsky2009, Bussieres2013} Of the rare-earth elements investigated for this application, erbium distinguishes itself as a natural choice for coupling directly to the low-loss optical fibers wiring the internet~\cite{Tittel2009} due to its $1.53~\,\mu\text{m}$ optical transition in the telecommunications C band with an optical coherence time of over 4 ms in yttrium orthosilicate crystals (YSO).~\cite{Boettger2003} For quantum memories based on atomic frequency combs or controlled reversible inhomogeneous broadening, initialization of the memory into a Zeeman state is inefficient in erbium doped YSO (Er$^{3+}$:YSO) compared to other rare earth ions because the optical lifetime (11 ms) is comparable to the Zeeman level lifetime ($\sim$100 ms).~\cite{Lauritzen2008} 
Detailed balance and parameters from Lauritzen et al.~\cite{Lauritzen2008} show spin initialization efficiency is limited to 68\% for a single pump beam.
Thus, while a few demonstrations of optical quantum memories based on Er$^{3+}$:YSO have already been reported,~\cite{Lauritzen2010} the efficiency of those memories were low and can be increased significantly by improving the memory initialization. Spin mixing using an RF source, or driving the spin-flip relaxation path with a second laser can be used to achieve an efficiency of over 90\%~\cite{Lauritzen2008}.  Alternatively, this lifetime reduction can also be accomplished by coupling the rare-earth atoms to on-chip microresonators with high quality factors and small optical mode volumes.~\cite{Zhong2015} A spin initialization of 90\% should also result from a reduction in the effective excited state lifetime by a factor of 6.
In this letter, we demonstrate photonic crystal resonators fabricated in Er$^{3+}$:YSO. The coupling of Er ions to the cavity results in enhanced interaction between the ions and photons coupled to the cavity mode. This allows for enhanced optical depth in a microns-long device and is thus an enabling technology for on-chip integrated telecom quantum memories. Excited state decay rate enhancement is demonstrated, which can be used to achieve higher efficiency optical pumping to a Zeeman level required to initialize the memory in protocols like atomic frequency combs.~\cite{Afzelius2009}

\begin{figure*}
\includegraphics{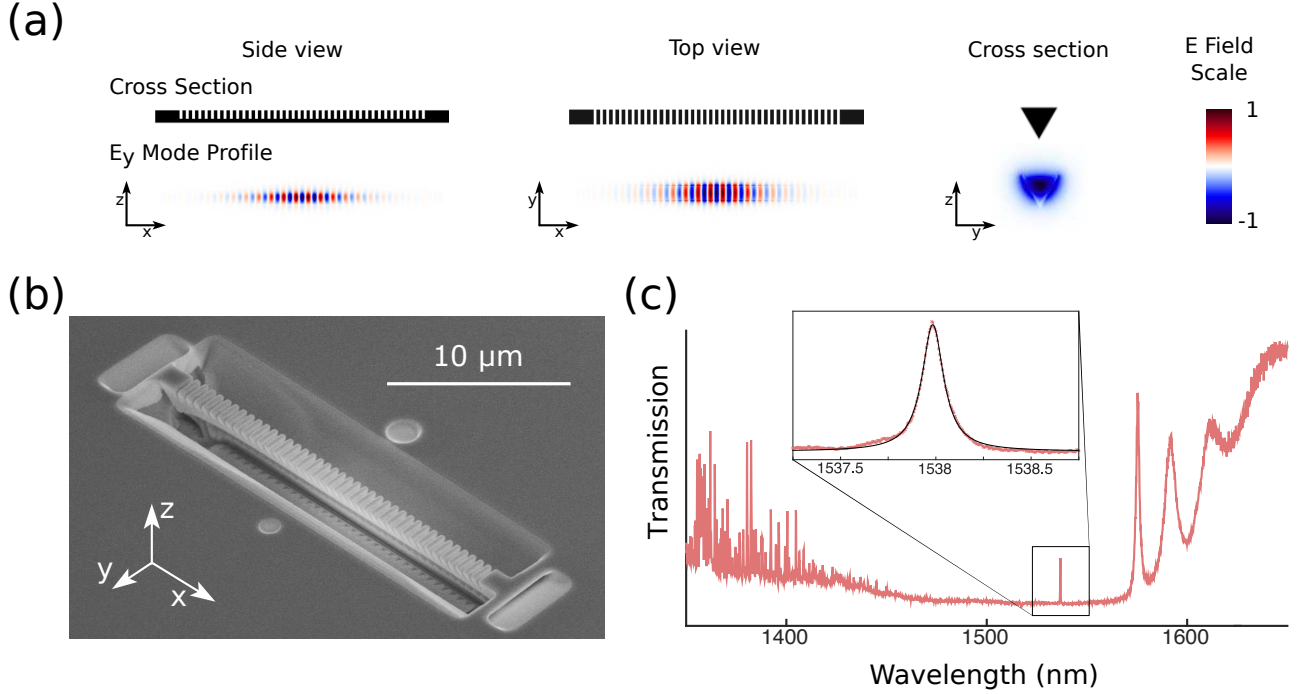}
\caption{(a) Cross sectional views of the triangular nanobeam through the center of the beam showing the structure of the beam (top) and the simulated cavity mode profiles (bottom). (b) Scanning electron microscope image of the triangular nanobeam YSO cavity. Angled trenches at the ends of the beam allow coupling from free space for transmission measurements. The x and y axes correspond to the optical axes \textbf{D$_2$} and \textbf{D$_1$}, respectively while z corresponds to the \textbf{b} axis of the orthorhombic YSO crystal. (c) Measured transmission through the nanobeam.  Broad-spectrum data taken with a supercontinuum laser; inset shows high-resolution frequency scan of a narrow linewidth laser in transmission through the cavity resonance at room temperature; fitting of a Lorentzian to the transmission spectrum shows the quality factor to be 11,400.  \label{beam}}
\end{figure*}

Triangular nanobeam optical cavities, like those demonstrated in our other work,~\cite{Zhongfab} were scaled to have a resonant wavelength matching the 1536 nm transition in Er$^{3+}$:YSO.  The cavity consists of an equilateral triangular beam with rectangular grooves milled into the top.  Each side of the triangle is $1.38~\,\mu\text{m}$ wide and the grooves are 200 nm wide and 800 nm deep with a 570 nm period.  The TE cavity mode possesses a simulated quality factor of Q$_{sim}$ = 70,000 and a mode volume of $V_{\text{mode}} = 1.65 (\lambda/n_{\text{YSO}})^3 = 1.05~\,\mu\text{m}^3$.  Here we have used the Purcell definition of mode volume.~\cite{Purcell1946} The cavity cross-section and optical mode profiles from finite difference time domain simulations using MEEP~\cite{MEEP} are shown in Fig.\ \ref{beam}(a).

The YSO crystal was grown by Scientific Materials Inc.\ with 0.02\% Er dopants.  The crystal was cut such that the \textbf{b} axis was normal to the polished top surface.  The resonator was milled using a focused ion beam system (FEI Nova 600).  The completed device is shown in Fig.\ \ref{beam}(b).  The nanobeam was oriented along the \textbf{D$_2$} direction of the YSO crystal, so that the electric field of the TE mode was oriented along the \textbf{D$_1$} direction of the crystal, where \textbf{D$_1$} and \textbf{D$_2$} are the optical axes of the biaxial birefringent YSO crystal.

\begin{figure}
\includegraphics{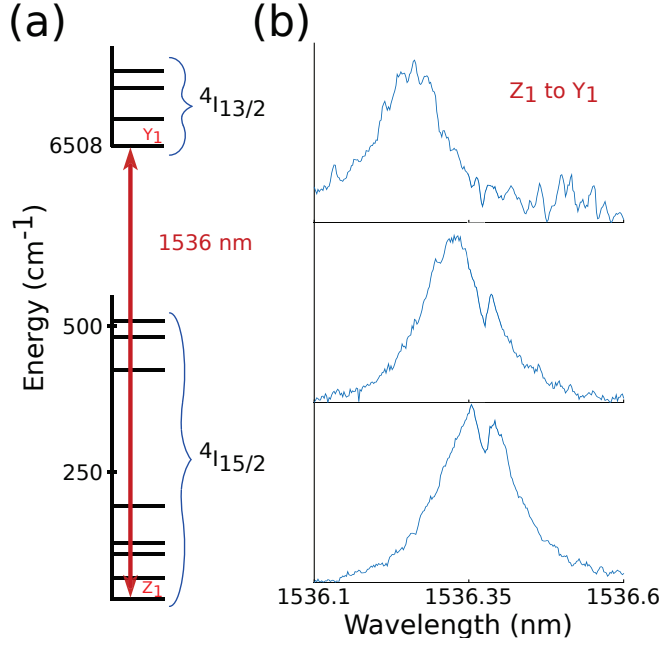}
\caption{(a) Erbium level diagram showing the crystal field splitting of the lowest and second lowest energy states. (b) Resonator transmission spectra as the cavity resonance was tuned using nitrogen deposition onto the 4.7K device.  The three steps show high resolution frequency scans as the cavity is tuned to the 1536 nm Er transition indicated in red. \label{tuning_and_levels}}
\end{figure}

The device was optically characterized using a custom-made confocal microscope.  The transmission spectrum was measured over a broad range of frequencies using a supercontinuum laser for input, with the output measured directly on a spectrometer with an InGaAs photodiode array (PyLoN IR 1024-1.7). The transmission through the beam is shown in Fig.\ \ref{beam}(c) showing the cavity resonance peak near 1536 nm. The quality factor of this resonance was determined to be Q=11,400 by least-squares fitting of a Lorentzian to the transmission curve. To attain higher resolution spectra, a tunable external cavity diode laser was stepped in frequency across the cavity, with the spectrometer integrating the transmission for 0.2 seconds per data point on the photodiode array. Upon cooling to 4.7K using a continuous flow liquid helium cryostat, the cavity resonance shifted to a higher frequency than the atomic resonance.  The atomic transition targeted was the 1536 nm transition between the lowest states of the $^4$I$_{15/2}$ and $^4$I$_{13/2}$ multiplets in the 4f orbital, labeled as Z$_1$ and Y$_1$ in Fig.\ \ref{tuning_and_levels}(a). This transition has an inhomogeneous broadening of $\sim$500MHz at liquid helium temperatures. Following the method presented in Mosor et al.,~\cite{Mosor2005} the optical resonance of the structure is precisely tuned to match the erbium absorption line, illustrated in Fig.\ \ref{tuning_and_levels}(b), by slowly letting nitrogen gas into the cryostat, which deposits onto the nanobeam.  The bottom scan on Fig.\ \ref{tuning_and_levels}(b) shows a dip that is 25\% the height of the full Lorentzian transmission peak.  As the power is lowered to reduce saturation, the size of the atomic absorption dip increases to $\sim$40\%.  The expected absorption coefficient in bulk for a field polarized along the \textbf{D$_1$} direction of the YSO crystal is 24.5 cm$^{-1}$.\cite{Bottger2006} A waveguide of the same length (26 microns) would have an attenuation of 3.8\%.  The resulting substantial increase in the optical depth is due to the interaction between the cavity mode and the Er ensemble.

\begin{figure}
\includegraphics{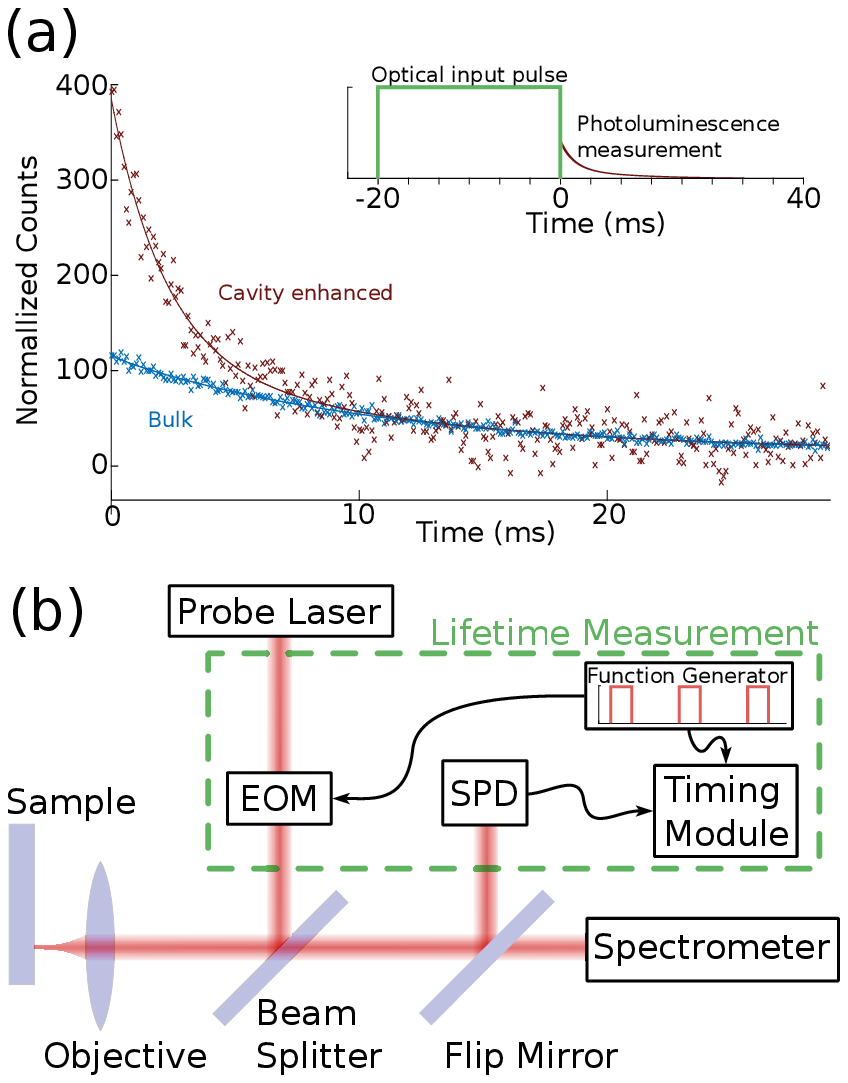}
\caption{(a) Photoluminescence decay from erbium ions as a function of time.  The cavity coupling decreases the lifetime via the Purcell effect.  The cavity-coupled luminescence converges to the bulk curve because all of the excited ions do not experience equal coupling. Inset shows the pulse used to excite the luminescence. (b) Simplified schematic of the confocal setup used to characterize the devices. With the flip mirror up, lifetime measurements were performed by modulating the input with an electro-optic modulator (EOM) synchronized with a single photon detector (SPD). \label{decay_and_confocal}}
\end{figure}

To measure the Purcell enhancement, the laser and resonator were tuned to the erbium transition line, and an electro-optic modulator was used to excite the ions with rectangular pulses 20 ms long with a 75 ms repetition period. The pulse and photoluminescence decay are shown in Fig. \ref{decay_and_confocal}(a) and the additions to the confocal microscope used for the lifetime measurement are shown in Fig. \ref{decay_and_confocal}(b).
 Time resolved photoluminescence measurements were taken with an IDQuantique ID220 InGaAs/InP avalanche photodiode detector. Given the measured quality factor and simulated mode volume, the expected Purcell enhancement for an ion positioned at the antinode of the cavity field is \begin{equation}
F_P = \frac{3}{4\pi^2}\left(\frac{\lambda}{n}\right)\left(\frac{Q}{V}\right) \left|\frac{E_{ion}}{E_{max}}\right|^2 = 517 
.
\end{equation}
Here $F_P$ is the Purcell factor, $\lambda$ is the cavity wavelength, $n$ is the cavity refractive index, $Q$ and $V$ are the cavity quality factor and mode volume, respectively.  The final term accounts for positional misalignment between the field and the dipole moment; $\vec E_{ion}$ is the electric field at the ion and $\vec E_{max}$ is the maximum of the electric field. 
Since the ensemble of ions is distributed uniformly inside the photonic crystal and the Purcell enhancement takes into account the emitter's dipole overlap with the field, most ions will not exhibit the full Purcell enhancement. The non-zero width of the inhomogeneous linewidth ($\sim$500MHz) was neglected, as it was much smaller than the cavity linewidth ($\sim$17GHz).  Taking these considerations into account~\cite{Zhong2015} gives an effective enhancement of 116.

Furthermore, the excited electrons can follow many decay paths from Y$_1$, of which only the path directly to Z$_1$ couples to the cavity mode and is thus enhanced. 
We estimate the branching ratio by comparing the expected emission rate, computed from the 1536 nm transition dipole moment, and the measurable 1/11.4 ms spontaneous decay rate.~\cite{McAuslan2009} 
%
For this calculation, we use the maximum absorption coefficient 24.5 cm$^{-1}$ with FWHM of 510 MHz for a 0.02\% erbium ion dopant density given an electric field polarized along the \textbf{D$_1$} direction from B\"ottger et al.\cite{Bottger2006} to compute an oscillator strength $f_{12} = 1.095 \times 10^{-7}$.  This is half the size of the value in McAuslan et al.~\cite{McAuslan2009} for \textbf{D$_2$} polarization due to the factor of two difference between absorption coefficient for light polarized along the \textbf{D$_1$} and \textbf{D$_2$} directions.
%
Following the results from McAuslan et al.,\cite{McAuslan2009} we find the spontaneous emission rate that we would expect from only this decay path to be 10.03 Hz. 
%
Comparing this value to the measured excited state decay rate of 87.7 Hz (11.4 ms lifetime), we determine that the branching ratio for Er:YSO in our cavity is $\sim$0.11. 
When taking this into account, the aforementioned factor of 116 increase in the spontaneous emission rate averaged over the cavity leads to a reduction in the excited state lifetime by a factor of 13, down to $\sim$900 $\mu$s. 

Fitting a single exponential, the lifetime in the bulk was found to be 10.8 ms, which is in agreement with values in the literature.~\cite{McAuslan2009} This was compared to the decay rate for ions in the cavity when the cavity is resonant with the ions.  In this case two exponential decays were fit, analogous to the fitting procedure by Y. Gong et al.~\cite{Gong2010}, and one of the decay curves had a time constant fixed at the bulk lifetime.  The bulk lifetime in this fit corresponds to ions that are not coupled to the optical cavity because they are located in the mirror sections.  The shorter lifetime was 1.8 ms. The luminescence data after the cavity had been tuned to be resonant with the ions is shown in comparison to bulk lifetime data in Fig.\ \ref{decay_and_confocal}(a).  The data was normalized by scaling the coefficient of the bulk decay rate.


Accounting for the branching ratio, the observed reduction in lifetime would correspond to an effective Purcell enhancement of $\sim$53. Due to the difficulty quantifying the number of excited ions per homogeneous linewidth, this analysis does not take into account the collective coupling effect, which could have contributed to the observed enhancement.  Future studies in this system will involve making cavities with higher quality factors, different dopant densities, and with the mode aligned to the \textbf{D$_1$} direction, which will allow a better assessment of the discrepancy between the expected reduction by a factor of 13 and measured reduction by a factor of 6.

In conclusion, we have fabricated an optical microresonator in an erbium-doped yttrium orthosilicate crystal, and used it to demonstrate enhanced optical depth and Purcell enhancement of the optical decay rate of the coupled erbium ions.
This is the first step to efficient on-chip solid-state quantum memories in the telecom C band. Next steps include the measurement of optical coherence of the cavity-coupled ions, as was demonstrated for the 883 nm transition of neodymium in YSO,~\cite{Zhong2015} and photon storage using the atomic frequency comb and controlled reversible inhomogeneous broadening techniques.


\begin{acknowledgments}
The authors sincerely thank Alexander E. Hartz for his contributions.  Financial support was provided an AFOSR Young Investigator Award (FA9550-15-1-0252), a Quantum Transduction MURI (FA9550-15-1-002), a National Science Foundation (NSF) CAREER award (1454607), and Caltech. Equipment funding was also provided by the Institute of Quantum Information and Matter (IQIM), an NSF Physics Frontiers Center (PHY-1125565) with support of the Gordon and Betty Moore Foundation (GBMF-12500028).  The device was fabricated in the Kavli Nanoscience Institute at Caltech with support from Gordon and Betty Moore Foundation.
\end{acknowledgments}




%

\end{document}